\documentclass[aps,prl,twocolumn,superscriptaddress]{revtex4}

\usepackage{booktabs}
\usepackage{multirow}
\usepackage{mathrsfs,amssymb,graphics,subfigure,threeparttable}
\usepackage{grap hicx}
\usepackage{amssymb}
\usepackage{enumerate}
\usepackage{amsmath}
\usepackage{fancyhdr}
\usepackage{bm}
\usepackage[squaren]{SIunits}
\usepackage{hyperref}
\usepackage{ulem}
\usepackage{textcomp}
\usepackage{color}

\begin{document}

\title{Efficient Generation of Tunable Magnetic and Optical Vortices Using Plasmas}

\author{Yipeng Wu}
\email{wuyipeng@ucla.edu}
\affiliation{University of California, Los Angeles, California 90095, USA}
\author{Xinlu Xu}
\affiliation{SLAC National Accelerator Laboratory, Stanford, California 94309, USA}
\author{Chaojie Zhang}
\affiliation{University of California, Los Angeles, California 90095, USA}
\author{Zan Nie}
\affiliation{University of California, Los Angeles, California 90095, USA}
\author{Mitchell Sinclair}
\affiliation{University of California, Los Angeles, California 90095, USA} 
\author{Audrey Farrell}
\affiliation{University of California, Los Angeles, California 90095, USA}
\author{Ken A Marsh}
\affiliation{University of California, Los Angeles, California 90095, USA}
\author{Jianfei Hua}
\affiliation{Department of Engineering Physics, Tsinghua University, Beijing 100084, China}
\author{Wei Lu}
\affiliation{Department of Engineering Physics, Tsinghua University, Beijing 100084, China}
\author{Warren B. Mori}
\affiliation{University of California, Los Angeles, California 90095, USA}
\author{Chan Joshi}
\email{cjoshi@ucla.edu}
\affiliation{University of California, Los Angeles, California 90095, USA}




\begin{abstract}

Plasma is an attractive medium for generating strong microscopic magnetic structures and tunable electromagnetic radiation with predictable topologies due to its extraordinary ability to sustain and manipulate high currents and strong fields. Here, using theory and simulations, we show efficient generation of multi-megagauss magnetic and tunable optical vortices when a sharp relativistic ionization front (IF) passes through a relatively long-wavelength Laguerre-Gaussian (LG) laser pulse with orbital angular momentum (OAM). The optical vortex is frequency upshifted within a wide spectral range simply by changing the plasma density and compressed in duration. The topological charges of both vortices can be manipulated by controlling the OAM mode of the incident LG laser and/or by controlling the topology and density of the IF. For relatively high (low) plasma densities, most energy of the incident LG laser pulse is converted to the magnetic (optical) vortex.

\end{abstract}


\maketitle

A recent paper proposed that a significant portion of the transverse electric field in an intense Laguerre Gaussian (LG) laser pulse can be transformed into a donut-shaped wake that has a significant longitudinal electric field 
used for
high-gradient particle acceleration in a plasma \cite{Vieira_PRL}.
This raises a complementary question: 
can we efficiently use the electromagnetic field of such a LG laser pulse to generate a helical current in a plasma and thereby excite a strong quasi-static and ultrashort-wavelength (micron-scale) magnetic vortex that has orbital angular momentum (OAM)?
In this letter we show using theory and three-dimensional (3D) particle-in-cell (PIC) simulations that the electromagnetic field of a relatively long-wavelength LG laser pulse can be efficiently 
transformed into a helical magnetic vortex structure by a counter propagating relativistic ionization front (IF). 
In addition, the topology of the magnetic field can be easily manipulated by controlling the topology of the IF
while the strength of the field can be varied by spatially changing the plasma density.
In addition, as it collides with the IF,
the LG laser pulse is frequency upshifted due to the relativistic Doppler effect and compressed since the number of optical cycles is conserved.
The resulting wavelength can be tuned over a wide spectral range by varying the plasma density of the IF. 
We also show that the 
OAM mode of the upshifted light can be manipulated by either changing the OAM mode of the incident LG pulse or by tuning the density and topology of the IF. 
Furthermore, a chirped LG pulse and static magnetic vortex of variable strength, both of which are difficult to obtain with state-of-the-art techniques, can be produced by propagating this IF in a density ramp.

Although much effort has been devoted to generating magnetic fields with different topologies in plasmas using intense laser pulses \cite{Haines_2001, Najmudin_2001, Ali_2010, Shi_2018, Nuter_2018, Sederberg_2020, Wilks_1992, Sudan_1993, Tatarakis_2002, Nilson_2006, Lampe_1978, Mori_1991, Fiuza_2010, Longman_PRR}, to our knowledge it is not yet possible to precisely control its azimuthal degree of freedom.
The present work is not only interesting from a fundamental perspective, 
but also important in areas ranging from condensed matter physics \cite{Blundell_2001} and accelerator/beam physics to nuclear fusion \cite{Post_1960} and astrophysics \cite{Mestel_1956}. 
For instance, azimuthal topological control will play an increasingly important role in the manipulation of electron spins \cite{Beaurepaire_1996} and topological quantum systems \cite{Edbert_2019, Zhang_2019}.
Such magnetic structures
could also
be used as ultracompact 
spiral undulators for generating coherent X-rays with OAM using electron beams with energy of just a few tens of MeV, 
as we have explored in Supplementary Material of this paper.

Similarly, OAM- and frequency-tunable radiation sources can be used in a diverse array of research disciplines, including optical manipulation and trapping \cite{optical_manipulation}, generation of high harmonics carrying OAM, super-resolution molecular spectroscopy, and fabrication of chiral organic materials \cite{Horikawa_15, Araki_18}.
We note that although 
the frequency upshifting of a non-vortex electromagnetic wave using a planar IF has been theoretically studied \cite{Lampe_1978, Mori_1991} and experimentally demonstrated in the microwave region \cite{IEEE_1, PRL_1992, IEEE_2},
these works did not involve OAM and topological control of the IF.

\begin{figure}[tp]
\includegraphics[height=0.26\textwidth]{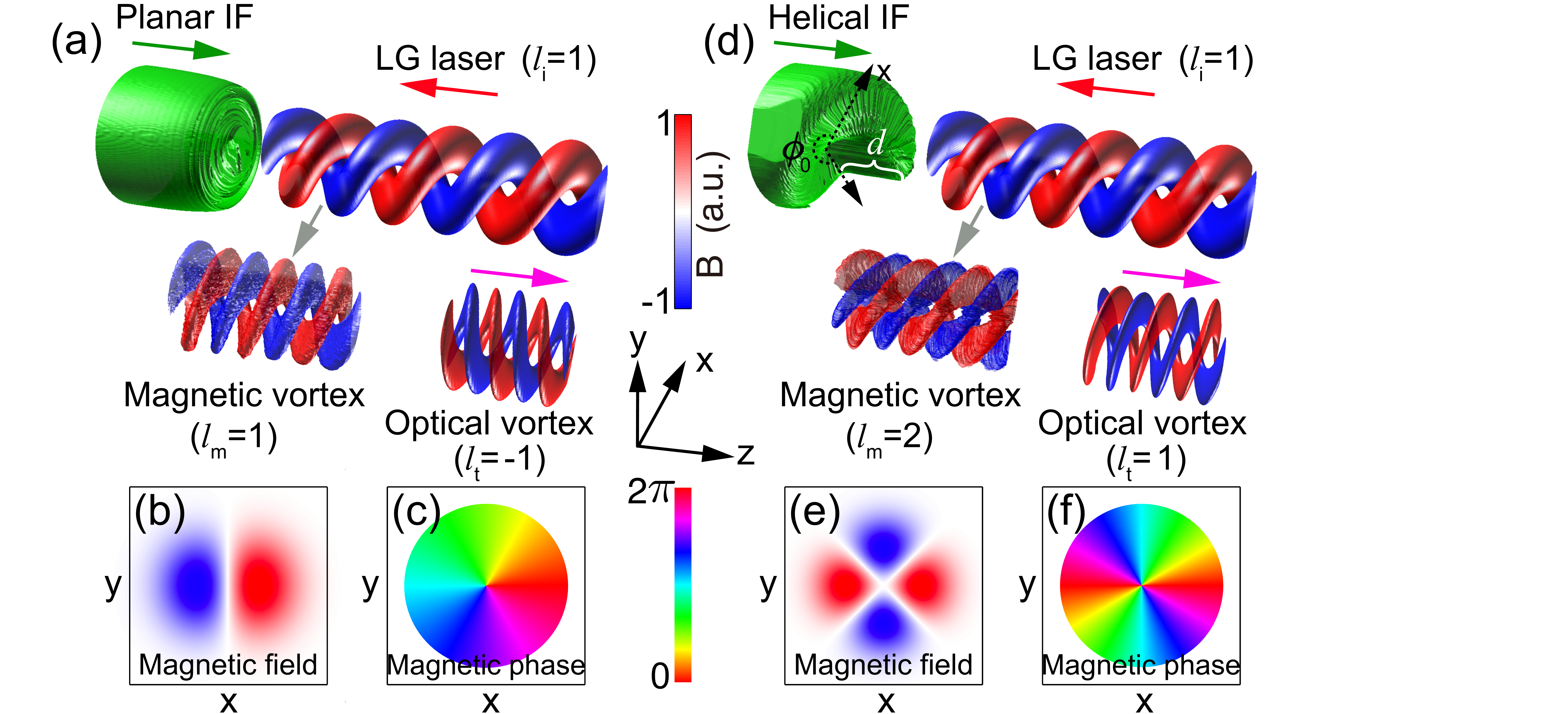}
\caption{\label{fig1} 
3D PIC simulations showing the generation of magnetic/optical vortex through the collision of a LG laser pulse with a planar (a) or helical (d) relativistic IF.
In both (a) and (d), the IF encounters and subsequently passes through the counter propagating LG mode laser pulse, producing a magnetic vortex and frequency upshifted optical vortex. The field distributions of the generated magnetic vortex at a particular $z$ are shown in (b) and (e) for the planar and helical IF case, respectively. The corresponding phase distributions of the magnetic vortex are shown in (c) and (f) for the planar and helical IF case, respectively.
}
\end{figure}

To gain a deeper insight into the generation of magnetic and optical vortices, we first carry out a theoretical analysis. 
The basic concept is schematically shown in Fig. 1. Here
a LG$_{p_il_i}$ laser with a wavelength of $\lambda_i$,
an azimuthal mode index (also called topological charge) of $l_i$, 
and zero radial mode index ($p_i=0$) is incident from the right on a planar [Fig. 1(a)] or helical [Fig. 1(d)] relativistic IF. The IF is generated via optical field ionization of neutral gas and moves from left to right with velocity $v_{f} \sim c$. 
Ahead of the IF is unionized gas, while behind the IF is stationary plasma.
Considering the interaction within the Rayleigh length and thus neglecting the Gouy phase effect,
the electric and magnetic fields of the incident LG laser have a phase term $E_i=cB_i\propto \exp(-ik_iz -il_i\phi - i\omega_i t)$ \cite{Allen_1992}, 
where $k_i=2\pi/\lambda_i$ is the wavenumber, 
$\phi=\arctan(y/x)$ is the azimuthal phase, $\omega_i=k_ic$ is the laser frequency
and $l_i$ gives rise to the OAM.
The ionizing laser used 
to produce the IF
is either an intense Gaussian laser pulse with a frequency of $\omega_{ion}\gg \omega_i$ (planar IF) or a spatiotemporally shaped light spring \cite{light_spring_OL, Vieira_PRL2} with a helical intensity profile consisting of co-propagating LG pulses with two different OAM modes differing by $\Delta l_{ion}$ and frequencies $\omega_{ion}\pm \Delta \omega_{ion}/2$ differing by $\Delta \omega_{ion}\ll  \omega_{ion}$ 
(helical IF). 
In the latter case, the light spring is 
of order $|\Delta l_{ion}|$,
consisting of $|\Delta l_{ion}|$ intertwined intensity helices 
with a temporal pitch of $|\Delta \tau_h|$, where $\Delta \tau_h=2\pi \Delta l_{ion}/\Delta \omega_{ion}$.
For simplicity, a first-order light spring with $|\Delta l_{ion}|=1$ is considered. 
When its intensity and rise time are properly chosen, it can ionize the neutral gas and generate a helical IF that has azimuthally varying thickness 
with a sudden jump at $\phi=\phi_0$ [see the green density iso-surface in Fig. 1(d)].
Without loss of generality, $\phi_0=2\pi$ is assumed and the longitudinal position of the IF can be expressed as $z_f(t, \phi)=v_f t+\frac{2\pi-\phi}{2\pi}d$.
Here, $d=0$ for the planar IF and $d=\alpha v_f \Delta \tau_h$ for the helical IF, with $\alpha$ a coefficient close to 1 (exact value mainly depending on the light spring's intensity and rise time). 
In the helical case, 
$z_f$ decreases (increases) in the anticlockwise direction from the value $v_f t+d$ at $\phi=0$ to the value $v_f t$ at $\phi=2\pi$ for $d>0$ ($d<0$).

In order to simplify the analysis, we work in the frame of the IF, which bears some similarity to Ref. \cite{Mori_1991}.
However, and most importantly, here we also consider the effects of both OAM and helical IFs. 
We denote quantities in the IF frame by primes. 
In this frame, the neutral gas flows into the IF from the right at velocity $v_f$, and plasma flows away to the left at the same velocity $v_f$. 
Through Lorentz transformation, 
we find that the IF is a stationary boundary (independent of $t'$) whose longitudinal position is 
$z_f'(\phi')=\frac{2\pi-\phi'}{2\pi}\gamma_fd$.
The incident wave is Doppler upshifted to $E_{i}'\propto \exp(-ik_{i}'z' -il_{i}'\phi' - i\omega_{i}' t')$, where $\omega_{i}'=(1+\beta_f)\gamma_f\omega_i$, $l_{i}'=l_i$ and $k_{i}'=\omega_{i}'/c$, while the plasma frequency is still $\omega_p=\sqrt{\frac{n_pe^2}{m_e\epsilon_0}}$ since it is Lorentz invariant. Here $\beta_f=v_f/c$ is the IF's velocity normalized to $c$, $\gamma_f=(1-\beta_f^2)^{-1/2}$ is the relativistic factor, 
$n_p$ is the plasma density in the laboratory frame, 
$e$ is the elementary charge, $m_e$ is the electron mass,
and $\epsilon_0$ is the vacuum electric permittivity.
To lowest order, the IF’s velocity is the group velocity of the ionizing laser in plasma. This gives $\beta_f\approx (1-\omega_p^2/\omega_{ion}^2)^{1/2}$ and $\gamma_f \approx \omega_{ion}/\omega_p$.

We first explain how the static magnetic vortex is generated for the helical IF. 
The corresponding physical quantities for the planar IF can be obtained simply by setting $d=0$.
In the IF frame, 
this magneto-static mode is a dispersionless electromagnetic wave, having frequency of $\omega_{m}'=\omega_{i}'$ and propagating at the plasma streaming velocity, i.e., $v_m'=v_f$. 
The corresponding wavenumber is given by $k_{m}'=\omega_{m}' /v_m'$.
Since the plasma has an azimuthally varying thickness for $d\neq0$, 
the phase front (equiphase surface) of this wave will undergo azimuthal modulation.
Assuming that at time $t_0'$, 
the phase front of such wave at $\phi'=2\pi$ is situated at $-v_m't_0'$, then the phase front position at all $\phi'$ is given by $z_0'(t_0',\phi')=-v_m't_0'+\frac{2\pi-\phi'}{2\pi}\gamma_fd\times(1-v_m'/c)-l_{i}'(\phi'-2\pi)/k_{m}'$ (Supplementary Note 1). 
For this mode with the form $E_{m}'=v_m' B_{m}' \propto \exp(-ik_{m}'z' -il_{m}'\phi' - i\omega_{m}' t')$, the phase front is the surface of the constant phase, determined by $k_{m}'z_0'(t_0',\phi') +l_{m}'\phi' +\omega_{m}' t_0'=constant$.
Therefore we can obtain 
$l_{m}'=l_{i}'+\frac{k_{m}' \gamma_f d}{2\pi}(1-v_m'/c)$. 
When transformed back to the laboratory frame this mode has $\omega_m=\gamma_f(\omega_{m}'-v_fk_{m}')=0$, $k_m=\gamma_f(k_{m}'-\omega_{m}'\beta_f/c)=k_i(1+1/\beta_f)\approx 2k_i$, $l_m=l_{m}'= l_{i}+\frac{d}{\beta_f \lambda_i }$, $E_m=0$, 
and $B_m \propto \exp(-ik_m z-l_m\phi)$. 
Therefore, in the laboratory frame this mode is a static magnetic vortex with a wavelength about half of the wavelength of the incident laser ($\lambda_m=2\pi/k_m\approx \lambda_i/2$) and a topological charge of $l_i+\frac{d}{\beta_f \lambda_i}\approx  l_i+\frac{d}{\lambda_i} $. The topological charge is mainly dependent on $l_i$ and $d$, and thus can be tuned by varying the incident OAM mode and IF's azimuthal thickness.

We also give an analysis in the laboratory frame to show the physical explanation of the origin of the magnetic vortex in Supplementary Note 2.
When the electrons are born (ionized) in the IF, they begin to move with velocity $v_e$ in the direction of the electric field of the incident laser. The high-frequency oscillation in the ionizing laser pulse averages out to zero over a few optical cycles.
The electric field of the low-frequency incident laser pulse gives rise to a periodic quasi-static current density $J=-en_pv_e$.
If the incident laser has a LG mode structure and/or the IF has a helical structure, the electron velocity $v_e$ and hence the electron current density $J$ has an azimuthal-dependence with $J\propto \exp(-ik_m z-l_m\phi+i\pi/2)$, which then generates an azimuthally periodic magnetic vortex
with the same magnetic polarization as the incident laser.

In addition to the magneto-static mode, a transmitted electromagnetic mode will also be generated when the incident laser interacts with the IF.
In the IF frame, this mode is a dispersive transmitted wave with frequency $\omega_{t}'=\omega_{i}'$, wavenumber $k_{t}'=\sqrt{\omega_{t}'^2-\omega_p^2}/c$, and phase velocity $v_{t}'=\omega_{t}'/k_{t}'$. 
Similarly, the transmitted phase front will undergo azimuthal modulation for the helical IF.
At time $t_0'$, 
the phase front position of such wave
at all $\phi'$ can be given by $z_0'(t_0',\phi')=-v_t't_0'+\frac{2\pi-\phi'}{2\pi}\gamma_fd\times(1-v_{t}'/c)-l_{i}'(\phi'-2\pi)/k_{t}'$ (Supplementary Note 3).
For the transmitted wave $E_{t}'\propto \exp(-ik_{t}'z' -il_{t}'\phi' - i\omega_{t}' t')$,
similar to the previous constant phase analysis, $z_0'(t_0',\phi')$ must meet the condition $k_{t}'z_0'(t_0',\phi') +l_{t}'\phi' +\omega_{t}' t_0'=constant$. 
Therefore 
 $l_{t}'=l_{i}'+\frac{k_{t}' \gamma_fd}{2\pi}(1-v_{t}'/c)$. 
Upon transforming back to the laboratory frame, the transmitted frequency and wavenumber can be obtained as $\omega_t=\gamma_f(\omega_{t}'-v_f k_{t}')$ and $k_t=\gamma_f(k_{t}'-\omega_{t}'\beta_f/c)$.
In the underdense limit where $\omega_p \ll \omega_{i}'$,
$\omega_t$ and $k_t$
can be approximated as $\omega_t \approx \omega_i(1+\frac{\omega_p^2}{4\omega_i^2}) \approx  \omega_i(1+\frac{n_p}{4n_{ic}})$ and $k_t \approx k_i(1-\frac{\omega_p^2}{4\omega_i^2}) \approx k_i(1-\frac{n_p}{4n_{ic}})$, with $n_{ic}$ the critical density of the incident laser.
When 
$n_p<4n_{ic}$,
$k_t>0$ and the transmitted wave moves forward in the laboratory frame with $E_t\propto \exp (-ik_{t}z -il_{t}\phi - i\omega_{t} t)$, where 
$l_t=l_{t}' \approx l_{i}-\frac{d}{\lambda_i}\times \frac{n_p}{4n_{ic}}$.
However, 
when $n_p>4n_{ic}$,
$k_t<0$ and the transmitted wave actually travels backwards in the laboratory frame (see Fig. 1) with $E_t\propto \exp (-i|k_{t}|z -il_{t}\phi + i\omega_{t} t)$, where $l_t=-l_{t}'\approx \frac{d}{\lambda_i}\times \frac{n_p}{4n_{ic}}-l_i$.
As one can see, 
the frequency-upshifted transmitted wave is also an optical vortex and its OAM mode depends on $l_i$, $d$ and $n_p$, which hence can be manipulated by tuning the incident OAM mode number, IF's azimuthal thickness, and plasma density.
Note that even for the planar IF ($d=0$), the transmitted OAM mode can also be tuned from $l_i$ to $-l_i$ by simply changing the plasma density from $n_p<4n_{ic}$ to $>4n_{ic}$.
In addition, for the helical IF ($d\neq0$), even though the incident laser is Gaussian mode without OAM ($l_i=0$), a transmitted laser with variable OAM can still be obtained.


\begin{table}[bp]  
\centering
\caption{\label{tab1} Summary of the key theoretical results.}   
\begin{tabular}{lcc}     
\hline
 & Magnetic vortex &Optical vortex \\    
 \hline
 Frequency &$\omega_m=0$ & $\omega_t \approx \omega_i(1+\frac{n_p}{4n_{ic}})$\\
Wavenumber  & $k_m \approx 2 k_i$ &$k_t\approx k_i(1-\frac{n_p}{4n_{ic}})$  \\  
Topological  &   \multirow{2}*{$l_m\approx l_{i}+\frac{d}{\lambda_i }$} & $l_t \approx l_{i}-\frac{d}{\lambda_i} \frac{n_p}{4n_{ic}}$ if $n_p<4n_{ic}$  \\ 
 charge& & $l_t \approx \frac{d}{\lambda_i} \frac{n_p}{4n_{ic}}-l_{i}$ if $n_p>4n_{ic}$\\
\hline
 \end{tabular}  
 \end{table}

\begin{figure}[bp]
\includegraphics[height=0.27\textwidth]{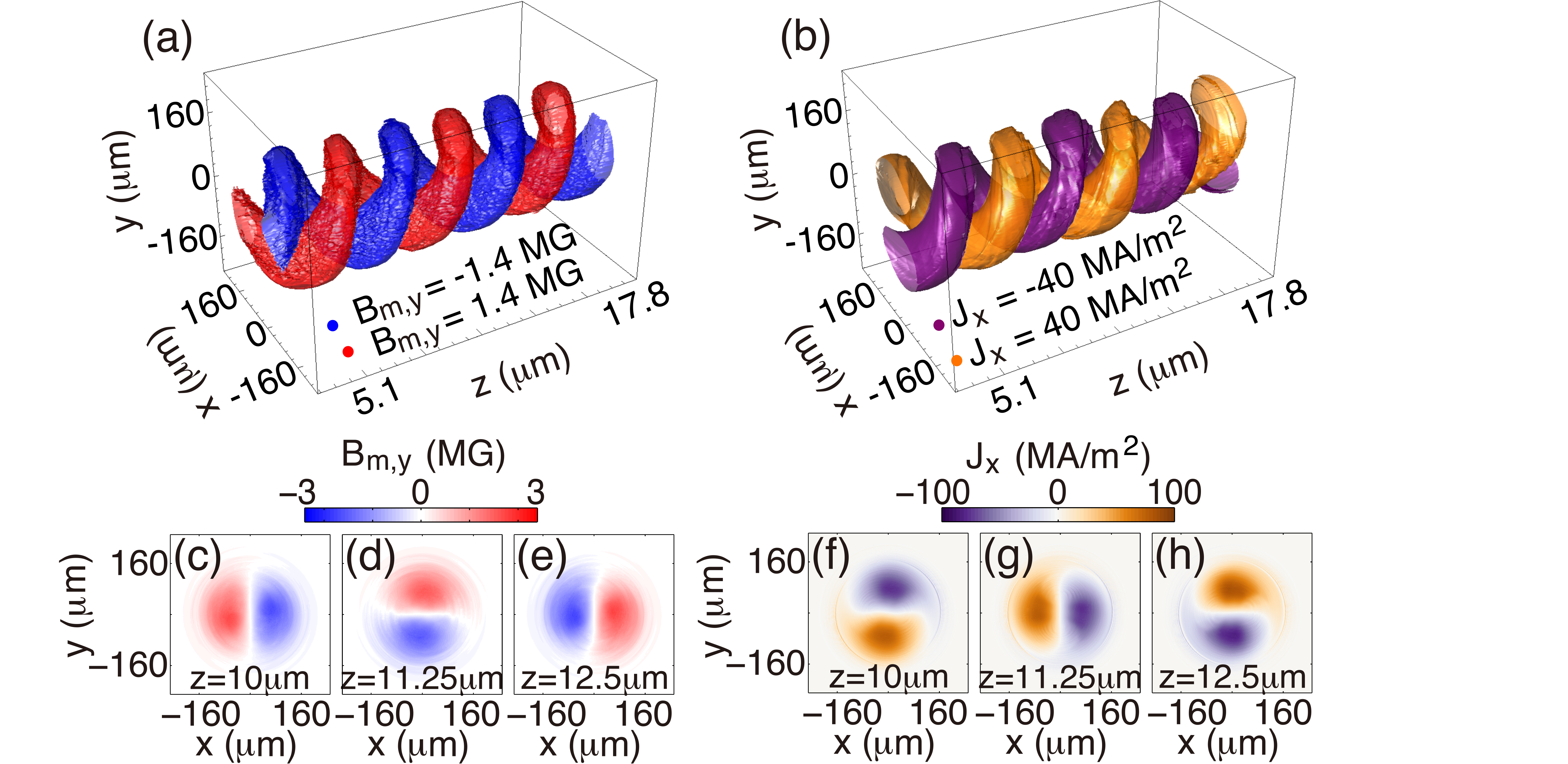}
\caption{\label{fig2} 
Simulation results of magnetic vortex generation through the collision of a linearly $x$-polarized LG$_{01}$ CO$_2$ laser ($l_i=1$) with a planar relativistic IF. 
The isosurfaces of generated magnetic field $B_{m,y}$ (a) and plasma current density $J_{x}$ (b). (c)-(e) The slices of $B_{m,y}$ at $z=$10 $\mu$m (c), 11.25 $\mu$m (d) and 12.5 $\mu$m (e). 
(f)-(h) The slices of $J_{x}$ at $z=$10 $\mu$m (f), 11.25 $\mu$m (g) and 12.5 $\mu$m (h). Note that every two adjacent slices are separated by $\lambda_m/4=1.25\ \mu$m.
}
\end{figure}


We also present the above theoretical analysis in matrix form in Supplementary Note 4 and summarize the key results in Table I.
To verify this analysis, we perform a series of 3D PIC simulations using the code OSIRIS \cite{Fonseca_2002}. 
In the planar IF case shown in Fig. 1(a), the IF is formed by the propagation of a linearly $y$-polarized laser pulse with a wavelength of 400 nm, FWHM duration of 20 fs, spot size on focus of 200 $\mu$m, 
and peak intensity of $7.5\times10^{15}$ W/cm$^2$ through a helium gas, which can ionize the outermost electron of helium atom.
The generated helium plasma has a density of $n_p = 1\times10^{20}$ cm$^{-3}$. 
The linearly $x$-polarized CO$_2$ LG$_{01}$ laser ($l_i=1$) has a wavelength $\lambda_i=10$ $\mu$m, a duration $\tau=0.5$ ps (flat-top temporal profile), a focused spot size $w$ = 100 $\mu$m, 
and a peak intensity $I_i \approx 5\times10^{14}$ W/cm$^2$, 
which is below the tunneling ionization threshold of the outermost electron of helium.
Since $n_p>4n_{ic}$ (here $n_{ic}= 1.11\times10^{19}$ cm$^{-3}$), we find that the transmitted wave moves backwards with a vacuum wavelength of 3 $\mu$m (in good agreement with the theoretical calculation $\lambda_t=2\pi c/\omega_t$) and an OAM mode number of $l_t=-1$ [see the optical vortex in Fig. 1(a)].

Figure 2 illustrates the 3D iso-surface distribution [Fig. 2(a)] and corresponding slices [Fig. 2(c)-(e)] of the generated magnetic vortex $B_{m,y}$, where the LG$_{01}$-like ($l_m=1$) vortex structure is clear. 
The magnetic field amplitude 
is quite large, about 2.85 MG (285 T). 
The current density $J_{x}$ supporting the magnetic field is shown in Fig. 2(b), which also features a helical iso-surface distribution. 
Both $J_{x}$ and $B_{m,y}$ have a wavelength $\lambda_m \approx \lambda_i/2\approx 5\ \mu$m. 
As expected from Ampere's law,
the phase difference of $B_{m,y}$ relative to $J_{x}$ is $\pi/2l_m$ (Supplementary Note 2), which can be readily seen from the comparison of Fig. 2(c)-(e) and Fig. 2(f)-(h).
If the incident laser polarization is changed from linear to circular, we find that the generated magnetic vortex in the plasma is also circularly polarized with the same handedness as that of the incident laser.
In addition,
such a magnetic vortex can last for a relatively long time ($\sim$10 ps).

\begin{figure}[bp]
\includegraphics[height=0.18\textwidth]{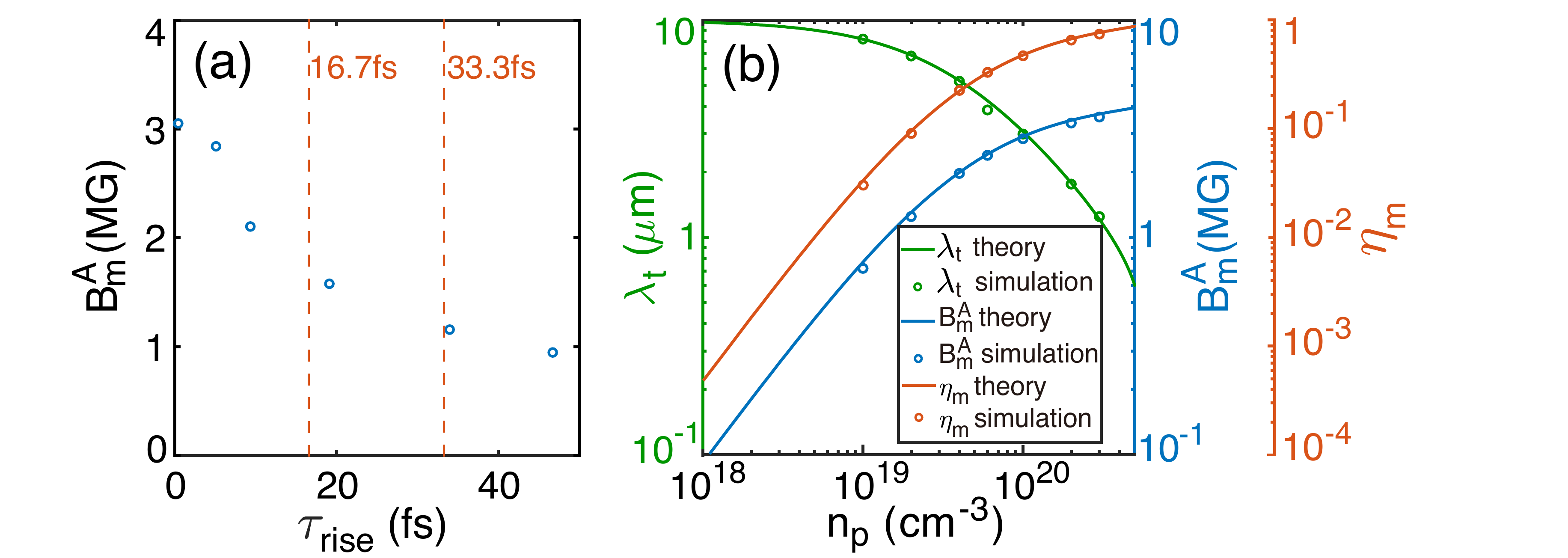}
\caption{\label{fig3}
(a) The amplitude $B_m^A$ of the magnetic vortex versus the rise time $\tau_{rise}$ of the IF for plasma density of $n_p = 1\times10^{20}$ cm$^{-3}$. (b) The transmitted vacuum wavelength $\lambda_t$, the amplitude $B_m^A$ and the energy conversion efficiency $\eta_m$ versus the plasma density $n_p$. 
}
\end{figure}

The above argument is valid if the rise time $\tau_{rise}$ of the IF is short compared to half the period $0.5\lambda_i/c$ of the incident CO$_2$ laser.
For the simulations in this paper, the helium gas is ionized in just $\sim$6 fs, which is much smaller than $0.5\lambda_i/c\approx16.7$ fs.
In the opposite extreme case shown in Fig. 3(a), if $\tau_{rise}>0.5\lambda_i/c$, the plasma electrons born in adjacent half cycles can have velocities in the opposite direction,
leading to a decrease in the plasma current and magnetic field amplitude.

In the case of a sharp IF, 
the transmission coefficient $\kappa_t$ and magnetic field coefficient $\kappa_m$ can be obtained, which are first calculated in the IF frame by matching electric and magnetic fields of these waves at the vacuum-plasma boundary and then transformed back to the laboratory frame \cite{Mori_1991}.
These two coefficients are given by $\kappa_t \equiv \frac{E_t^A}{E_i^A}=\frac{2}{1+\sqrt{\delta}}$ and $\kappa_m \equiv \frac{B_{m}^A}{B_{i}^A}=2\beta_f\frac{1-\sqrt{\delta}}{1-\beta_f\sqrt{\delta}}$,
where $E_t^A$, $E_i^A$, $B_{m}^A$, and $B_{i}^A$ are the amplitudes of the transmitted electric field, incident electric field, generated static magnetic field and incident magnetic field in the laboratory frame, respectively, 
and $\delta=1-n_p/[(1+\beta_f)^2\gamma_f^2n_{ic}]$.
In the underdense limit, $\kappa_t$ approaches unity and the energy conversion efficiency from the incident laser to the transmitted wave 
is $\eta_t \approx \lambda_t/\lambda_i$, assuming that the IF 
moves across the entire duration of the incident laser.
With the same assumption, the energy conversion efficiency from the incident laser to the static magnetic field is $\eta_m \approx \kappa_m^2/4$.
As one can see, all these quantities are dependent on 
the plasma density $n_p$.
We have run 3D OSIRIS simulations to show the dependence of $\lambda_t$, $B_{m}^A\equiv \kappa_m B_{i}^A$ and $\eta_m$ on $n_p$ using the CO$_2$ laser parameters corresponding to Fig. 2 ($B_{i}^A\approx2.05$ MG), as shown in Fig. 3(b).
Good agreement between theory and simulation has been achieved. 
The transmitted vacuum wavelength $\lambda_t$ decreases with an increase in $n_p$ and can be tuned in the spectral range of 1-10 $\mu$m by simply tuning $n_p$.
For high densities, substantial wavelength downshifts (frequency upshifts) occur.
Meanwhile, 
$\kappa_m$ approaches 1.9, $B_{m}^A$ approaches 3.9 MG and $\eta_m$ approaches 0.9, 
meaning that 90$\%$ incident CO$_2$ laser energy remains stored in the magnetic vortex.

\begin{figure}[bp]
\includegraphics[height=0.25\textwidth]{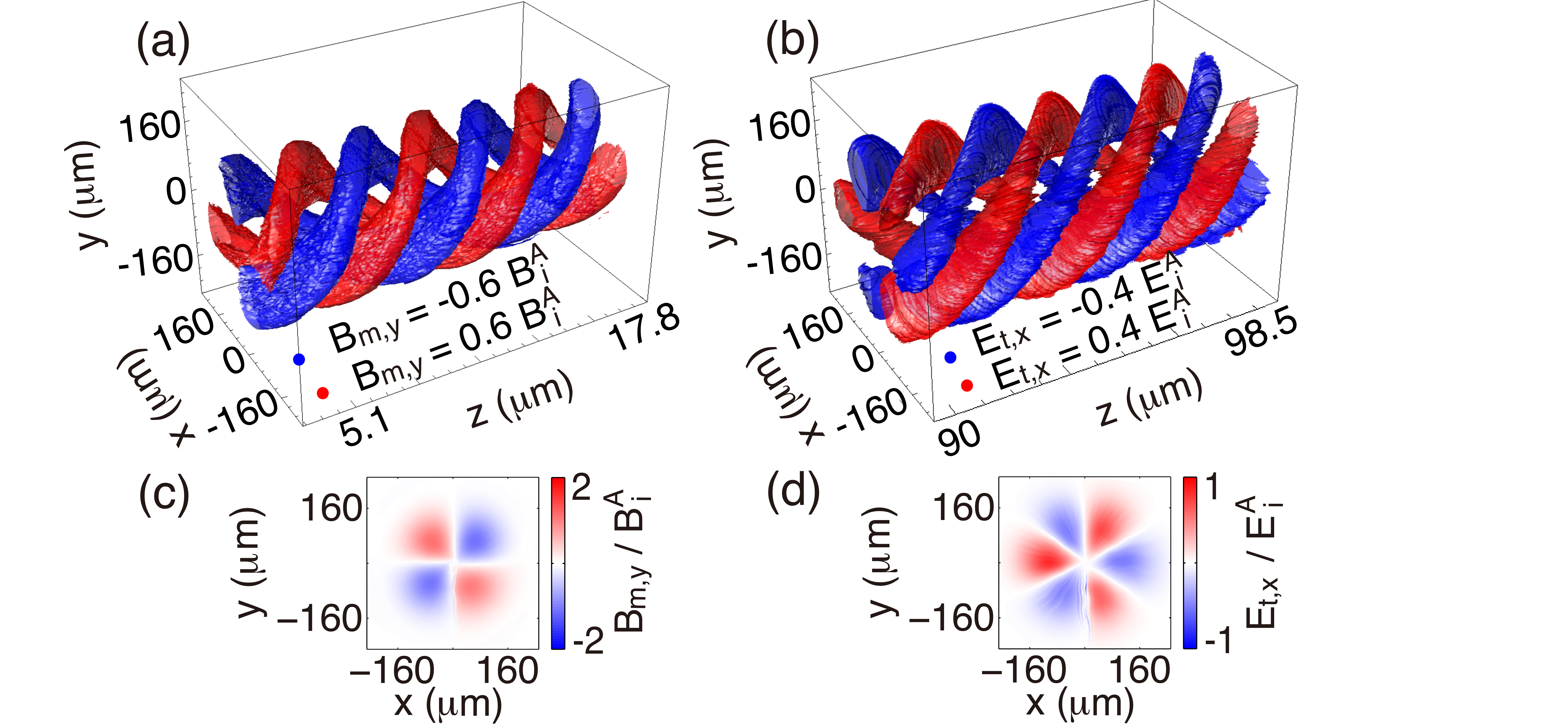}
\caption{\label{fig4}
Simulation results of magnetic/optical vortex generation through the collision of a linearly $x$-polarized LG laser with a helical IF. 
(a) and (c) show the isosurface and corresponding slice of the generated magnetic vortex $B_{m,y}$ (normalized to the incident magnetic amplitude $B_i^A$) with the incident OAM mode $l_i=1$. 
(b) and (d) show the isosurface and corresponding slice of the transmitted optical vortex $E_{t,x}$ (normalized to the incident electric amplitude $E_i^A$) with the incident OAM mode $l_i=-1$. 
}
\end{figure}

Next, we present the simulation results for the helical IF.
In these simulations, a first-order light spring composed of dual-frequency 
LG$_{01}$ (wavelength 406.5 nm) and LG$_{02}$ (wavelength 393.7 nm) modes 
is used to ionize helium and generate a helical IF with $d= \lambda_i$ and $n_p=8n_{ic}=8.9\times10^{19}$cm$^{-3}$.
When the incident CO$_2$ laser has an OAM mode number of $l_i=1$, the generated magnetic vortex features a double-twisted LG$_{02}$-like helical structure with $l_m \approx l_i+\frac{d}{ \lambda_i }=2$ [see Fig. 4(a) and 4(c)] and the transmitted optical vortex moves backwards with an OAM mode number of $l_t\approx \frac{d}{\lambda_i}\times \frac{n_p}{4n_{ic}}-l_i=1$ [see the optical vortex in Fig. 1(d)].
If $l_i$ is changed from 1 to -1, the generated magnetic field structure will have a planar topology with $l_m=0$. However, the OAM mode number of the transmitted optical vortex will be increased to $l_t\approx \frac{d}{\lambda_i}\times \frac{n_p}{4n_{ic}}-l_i=3$, as shown in Fig. 4(b) and 4(d), where a triple-twisted LG$_{03}$-like helical structure can be clearly seen.
Furthermore, by tuning $d$ and/or $n_p$, non-integer $l_m$ and $l_t$ values can also be obtained, where the magnetic and optical vortices have complicated phase structures and unique features \cite{Berry_2004} (see Supplementary Note 5 for examples of non-integer magnetic vortices with $l_m=1.25$ and $l_m=1.5$).

Such ultra-short-wavelength, perfectly periodic, bi-directional, and long-lifetime magnetic vortices can be used as ultra-compact spiral undulators for the production of high-power superradiant X-ray radiation with OAM.
Taking the generated LG$_{01}$-like ($l_m=1$) magnetic vortex with a wavelength of $\lambda_m=5$ $\mu$m and an amplitude of $B_m^A=3.5$ MG as an example, 
simulations show that 1-nm, coherent vortex X-ray radiation with power of $\sim$1 MW could be emitted from such a spiral undulator of just $\sim$1 cm in length by using  a 68-pC, 25-MeV, moderately pre-bunched electron beam.
The OAM mode number of the radiation $l_r$ is opposite of $l_m$, i.e., $l_r=-l_m=-1$, and therefore can be further manipulated by tuning $l_m$ (Supplementary Note 6).

It should be noted that undulator shaping is preferred for optimizing the radiation performance.
In addition to being a magnetic ``convertor", the plasma can also serve as a magnetic ``shaper" to shape both the longitudinal and transverse amplitude distributions 
by tailoring the plasma density structure due to the amplitude's strong dependence on the local plasma density (see Supplementary Note 7 for an example of longitudinal tailoring).
This makes the spiral undulator proposed here advantageous over an optical undulator, since it is very difficult to directly tailor the laser itself even with state-of-the-art techniques.

In summary, we have proposed a novel scheme that can efficiently generate multi-megagauss magnetic vortices and tunable optical vortices. This scheme is based on the interaction of a LG laser pulse with a tailored relativistic IF. Such magnetic and optical vortices have the potential to be used in numerous applications, such as generation of coherent X-rays with OAM.

\section{Acknowledgements}
We are grateful to Dr. Frederico Fiúza. This work was supported by the U.S. Department of Energy Grant No. DE-SC0010064 and NSF Grant No. 1734315. The simulations were performed on the NERSC Cori cluster at LBNL.

\section{references}


\end{document}